\documentclass[runningheads]{llncs}
\usepackage[T1]{fontenc}
\usepackage{graphicx}
\usepackage{comment}
\usepackage{layouts}
\begin{document}
\title{MRI-based Head and Neck Tumor Segmentation Using nnU-Net with 15-fold Cross-Validation Ensemble}

\titlerunning{MRI-based HNC segmentation by 15-fold cross-validation ensemble}

\author{
Frank N. Mol\inst{1}\orcidID{0000-0002-4342-1126} \and
Luuk van der Hoek\inst{2}\orcidID{0009-0006-9866-4527} \and
Baoqiang Ma\inst{2}\orcidID{0000-0002-6461-1244} \and
Bharath Chowdhary Nagam\inst{1}\orcidID{0000-0002-3724-7694} \and
Nanna M. Sijtsema\inst{2}\orcidID{0000-0001-6644-274X} \and
Lisanne V. van Dijk\inst{2}\orcidID{0000-0002-9515-5616} \and
Kerstin Bunte\inst{1}\orcidID{0000-0002-2930-6172} \and
Rifka Vlijm\inst{1}\orcidID{0000-0001-8909-9518} \and
Peter M.A. van Ooijen\inst{2}\orcidID{0000-0002-8995-1210} }

\authorrunning{F. N. Mol et al.}

\institute{University of Groningen, Nijenborgh 4, 9747 AG Groningen, The Netherlands\\
\email{\{frank.mol, b.c.nagam, k.bunte, r.vlijm\}@rug.nl}\and
University Medical Center Groningen (UMCG), 9700 RB Groningen, The Netherlands
\email{\{l.van.der.hoek02, b.ma, n.m.sijtsema, l.v.van.dijk, p.m.a.van.ooijen\}@umcg.nl}}

\maketitle

\begin{abstract}
The superior soft tissue differentiation provided by MRI may enable more accurate tumor segmentation compared to CT and PET, potentially enhancing adaptive radiotherapy treatment planning.
The Head and Neck Tumor Segmentation for MR-Guided Applications challenge (HNTSMRG-24) comprises two tasks: segmentation of primary gross tumor volume (GTVp) and metastatic lymph nodes (GTVn) on T2-weighted MRI volumes obtained at (1) pre-radiotherapy (pre-RT) and (2) mid-radiotherapy (mid-RT). 
The training dataset consists of data from 150 patients, including  MRI volumes of pre-RT, mid-RT, and pre-RT registered to the corresponding mid-RT volumes. 
Each MRI volume is accompanied by a label mask, generated by merging independent annotations from a minimum of three experts.
For both tasks, we propose adopting the nnU-Net V2 framework by the use of a 15-fold cross-validation ensemble instead of the standard number of 5 folds for increased robustness and variability.
For pre-RT segmentation, we augmented the initial training data (150 pre-RT volumes and masks) with the corresponding mid-RT data. 
For mid-RT segmentation, we opted for a three-channel input, which, in addition to the mid-RT MRI volume, comprises the registered pre-RT MRI volume and the corresponding mask.
The mean of the aggregated Dice Similarity Coefficient for GTVp and GTVn is computed on a blind test set and determines the quality of the proposed methods.
These metrics determine the final ranking of methods for both tasks separately.
The final blind testing (50 patients) of the methods proposed by our team, $RUG\_UMCG$, resulted in an aggregated Dice Similarity Coefficient of 0.81 (0.77 for GTVp and 0.85 for GTVn) 
for Task 1 and 0.70 (0.54 for GTVp and 0.86 for GTVn) 
for Task 2.

\end{abstract}
\section{Introduction}
Head and neck cancer (HNC) is among the most 
 prevalent cancers globally, with over half a million new cases each year, making it the eighth leading cause of cancer-related deaths \cite{sung2021global}. 
 In clinical practice, HNC patients are often treated with (chemo-)radiotherapy, optionally complemented with surgery.
  Accurate tumor segmentation is essential for precisely targeting tumor regions with the appropriate dose while sparing surrounding healthy tissues.
 The tumor contours are generally manually delineated by experts on CT scans, using MRI and PET scans as references.
However, MRI offers superior soft tissue differentiation compared to CT and PET, and MRI-only data facilitates adaptive RT with MR-linac.
 The manual delineation is time-consuming and may introduce intra- and inter-observer variability \cite{Valentini2014}.
 Automatic tumor segmentation could tackle 
 these challenges by advanced techniques such as deep learning.
 %
Similar challenges preceding HNTSMRG-24 predominantly explored automated segmentation of HNC on CT and PET scans, as in the HECKTOR challenge of 2020, 2021, and 2022 \cite{10.1007/978-3-031-27420-6_1,andrearczyk2021overview,andrearczyk2020overview}.

Previous studies primarily utilized Deep Convolutional Neural Networks\\ 
(DCNNs) for segmenting head and neck tumors in PET/CT and MRI scans \cite{dcnn_2,dcnn_1,dcnn_3,WAHID20226}.
Despite being introduced almost a decade ago, the U-Net architecture \cite{Unet} continues to be dominant in the field of medical image segmentation.
The (original) 2D variant gained prominence in segmenting the gross tumor volume (GTV) \cite{2dunet_2}.
Further developments have been made in the form of 2.5D (handling multiple slices at once) and 3D (handling complete volumes) \cite{3dunet}.
Due to the increasing computation capabilities, the 3D U-Net has widely been adopted for head and neck tumor segmentation, yielding impressive results across multiple studies \cite{3dunet_3,3dunet_2,3dunet_1}. Furthermore, a comparison of the different U-Net approaches resulted in the best segmentation performance for 3D U-Net in MRI segmentation, at the cost of increased computation time \cite{unet_dimension_comparison}.
The tuning of hyper-parameters and design choices of such a U-Net have also been automated for medical image segmentation by the nnU-Net framework \cite{nnUnet}, which recently received a major update \cite{nnunetv2}. This framework is already widely used in the HNC tumor segmentation for PET and CT modalities \cite{nnUnet_2,nnUnet_1}.

Alternatively to the DCNN paradigm, transformer-based models have recently garnered significant attention in the fields of image processing and computer vision, and consequently the field of medical image segmentation.
The Swin UNETR \cite{swinunetr} has emerged as a popular architecture, demonstrating remarkable success in 3D brain tumor segmentation. 
Meta introduced the Segment Anything Model (SAM) \cite{SAM} as a foundation model for zero-shot image segmentation, marking a significant advancement in the field.
Despite being pretrained on approximately 11 million natural images, SAM lacks performance in medical applications due to absence of medical images in training.
MedSAM \cite{medsam} was developed to address this issue, by utilizing the SAM model and pretraining it on over a million annotated medical images from more than 30 cancer types.
MedSAM has been applied to head and neck tumor segmentation in various recent studies \cite{medsam_1}, demonstrating its potential in specialized medical imaging tasks.

MRI data offers significant potential for HNC tumor segmentation compared to CT and PET, due to its superior soft tissue contrast provided by multi-sequence settings.
Among the commonly employed MRI techniques, including T1-, T2-weighted, and diffusion-weighted imaging (DWI), T2-weighted MRI scans are often favored, in case of single-modality use, for tumor segmentation due to their enhanced sensitivity to water content. 
This enhanced contrast between tumors and surrounding tissues facilitates more accurate delineation.
%
%
In radiotherapy, MRI scans typically precede the start of treatment. 
However, with the emergence adaptive radiotherapy has led to the widespread use of MRI scans during the course of the treatment.
The data considered in this study consists of two scans for every patient. A pre-radiotherapy (pre-RT) scan acquired 1-3 weeks before treatment and a mid-radiotherapy (mid-RT) scan acquired 2-4 weeks after the first treatment. These scans are complemented  with expert annotations identifying primary gross tumor volumes (GTVp) and metastatic lymph nodes (GTVn) and hence capture changes in these gross tumor volumes for each patient. 

This study aims to develop a deep learning model for HNC tumor segmentation based on T2 MRI volumes by comparing a custom Unet, a model trained using nnU-Net V2 and SAM. To increase robustness and variability, the nnU-Net 5-fold cross validation ensemble was adjusted to a 15-fold cross validation ensemble. The performance of the networks was optimized for two tasks: GTVp and GTVn segmentation on either (1) pre-RT or (2) mid-RT MRIs and was assessed by the aggregated Dice Similarity Coefficient with the ground truth segmentations.

\section{Materials and Methods}
\subsection{Data}
The training dataset comprises data from  %
150 patients who were treated for HNC with (chemo)radiation at
The University of Texas MD Anderson Cancer Center (MDACC).
For each patient, the T2-weighted MRI scans of pre-RT (1-3 weeks before start of RT) and mid-RT (2-4 weeks after start of RT) were provided.
Furthermore, registered pre-RT volumes were provided (registered on the corresponding mid-RT MRI volume), for which the SimpleElastix package was used \cite{simpleelastix}.
The data were provided in NIfTI format, with a uniform spacing of $0.5 \times 0.5 \times 2.0$ mm applied across all datasets.
Each volume is a cropped MRI containing the region from the top of the clavicles to the bottom of the nasal septum.
The sizes of the volumes vary across patients. E.g. the number of slices in the pre-RT training volumes varies between 57 and 162. 
To ensure uniformity in shape for the training samples, this variability needs to be adjusted.%

\begin{figure}[t]
\includegraphics[width=\textwidth]{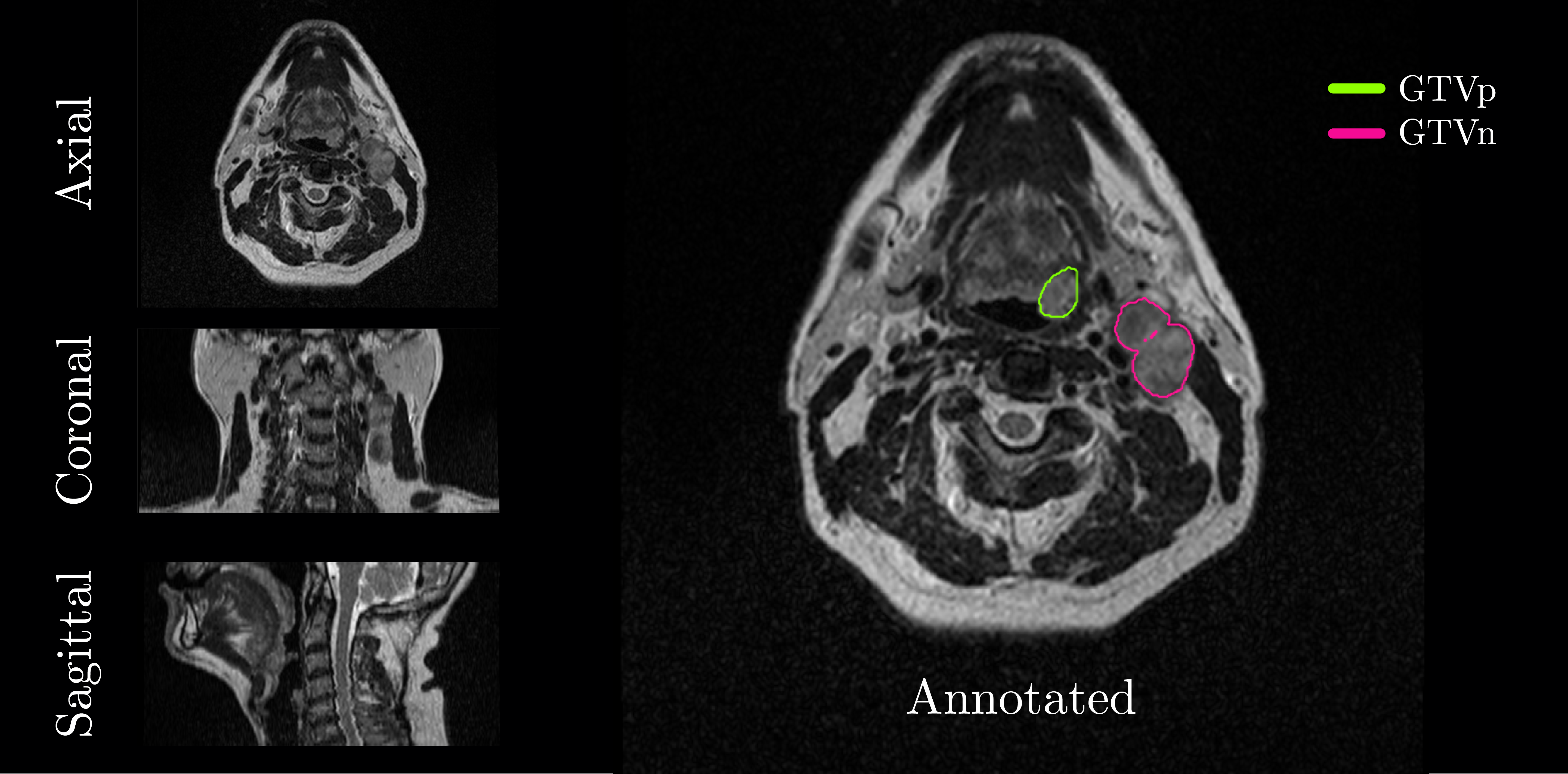}
\caption{Example of pre-RT MRI volume with corresponding annotations. The left panel shows the axial, coronal and sagittal view of the center of volume. The right panel displays the annotated labels for GTVn and GTVp in the axial view. The displayed data is from patient number 30 from the training set.}
\label{fig1}
\end{figure}

For each MRI volume, a corresponding label mask was provided, containing a voxel-wise annotation for three classes: background, GTVp, and GTVn (class 0, 1, and 2, respectively).
These annotations were based on a combination of manual annotations from at least three experts.
%
Fig. \ref{fig1} shows the pre-RT MRI sample of patient number 30 with corresponding annotation.

The GTVp volumes in the pre-RT training data have a volume consisting of on average of 22,138 voxels (137 GTVp volumes across 135 patients). In the mid-RT data, the average number of voxels for GTVp significantly dropped to 8,373 (116 GTVp volumes across 114 patients). For GTVn, the average number of voxels per volume had a lower reduction: 14,001 (259 voxels, with 20 patients having no GTVn) for pre-RT to 9,607 for mid-RT (241 voxels, with 21 patients not having GTVn).

During development, the used methods were tested using a Docker image on a preliminary blind test set consisting of data from two patients for both Task 1 and Task 2.
Similarly, the final test is conducted blindly on a test set comprising data from 50 patients (not present in the initial training data).
%
For Task 1, a pre-RT MRI volume was provided per patient as a single file as input for testing.
%
For Task 2, five files per patient were provided for input for testing: (1) the mid-RT MRI volume, (2) the corresponding pre-RT MRI volume and (3) its respective label mask, (4) the registered pre-RT MRI volume and (5) its respective label mask.

\subsection{Preprocessing}
The preprocessing steps were determined by the nnU-Net v2 framework by automatically analyzing the training data, resulting in similar preprocessing for both tasks. 
The main step of this preprocessing is Z-score normalization of the MRI volumes before training (and for inference). 
The original voxel size of $0.5 \times 0.5 \times 2.0$ mm$^3$ was kept, i.e. no resizing of the volumes was performed.
Since the data consists of large, variably sized volumes, a crop of the volumes was used during training, for example, crops of shape (48, 224, 192) were used in Task 1.
These crops were taken randomly, with oversampling the foreground labels (GTVp and GTVn) using a probability of $0.33$. 
The oversampling was used to overcome the class imbalance in training (e.g. $99.8\%$ of the voxels is background).
During training, the data was augmented following the recommended augmentations of the nnU-Net V2 framework \cite{nnUnet}, e.g. rotations, mirroring, blur.
For inference, these volumes are again cropped and fed forward to the network using a sliding-window approach with a step size of 0.5 (the cropped volumes have 50 \% overlap).

\subsection{Architecture}
\begin{figure}[t]
\includegraphics[width=\textwidth]{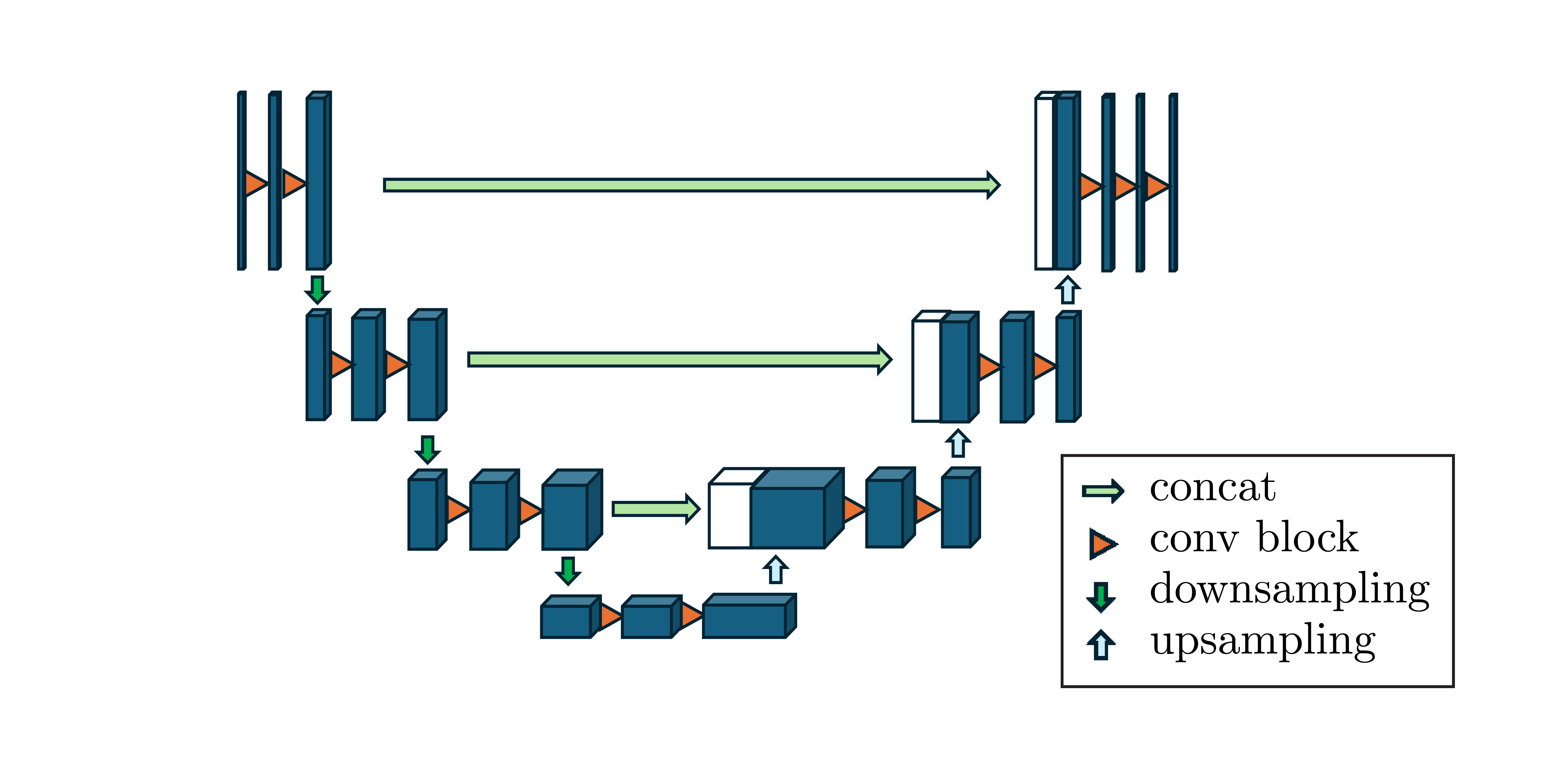}
\caption{Schematic of the U-Net architecture, as conceptually proposed by Ronneberger et al. \cite{Unet}. In our method, we use the volumetric U-Net that processes 3D volumes.}
\label{fig2}
\end{figure}
The Deep-Learning architecture used for both tasks was a 3D U-Net \cite{3dunet,Unet}. In Fig. \ref{fig2} a schematic of the general U-Net architecture is presented. We have used the nnU-Net V2 framework \cite{nnUnet,nnunetv2} for planning and training the models. We have adopted the \textit{3d\_fullres} configuration due to the proven performance of this configuration \cite{unet_dimension_comparison}, the framework also provides planning for \textit{2d}, \textit{3d\_lowres} and \textit{3d\_cascade\_fullres}.
LeakyReLU activation functions were used with a negative slope set to $\alpha = 0.01$. The loss function that was used is a combination of the Dice Loss and Cross Entropy loss (sum with equal weights). Stochastic Gradient Descent was used as the optimizer, using a learning rate scheduled from the initial $0.01$ to $10^{-5}$ over the used epochs.

\subsection{Evaluation}
The evaluation of the segmentation quality of the proposed methods is done by computing the aggregated  Dice Similarity Coefficient ($DSC_{agg}$), obtained from \cite{dice_agg}. This score is defined by
\begin{equation}
    DSC_{agg} = \frac{2 \times \sum_{k} TP_{k}}{2 \times \sum_{k} TP_{k} + \sum_{k} FP_{k}  +\sum_{k} FN_{k}} = \frac{2\sum_{k}\bar{y_{k}}y_{k}}{\sum_{k}(\bar{y_{k}}+y_{k})}
\end{equation}
with $k$ as the patient number, $TP$, $FP$ and $FN$ are the true positives, false positives and false negatives, respectively, and $\bar{y_{k}}$ is the prediction for the ground truth volume $y_{k}$ for patient number $k$.
This score is an extended version of the original Dice Similarity Coefficient, which is given by
\begin{equation}
    DSC = \frac{2 \times TP_{k}}{2 \times TP_{k} +  FP_{k}  + FN_{k}} = \frac{2\bar{y_{k}}y_{k}}{(\bar{y_{k}}+y_{k})}
\end{equation}
for an individual patient $k$.
Typically, to assess the overall performance of a method on an entire test set, the mean $DSC$ is calculated across all patients per type of GTV volume.
A disadvantage of this metric is that test samples with small GTV volumes can yield a low individual $DSC$, as even a small number of incorrect voxels can significantly affect the final result. 
The $DSC_{agg}$ overcomes this problem by calculating the final metric voxel-wise alternatively to sample-wise, which reduces the the large impact of small GTV volumes.
\begin{figure}[t]
\centering
\includegraphics[width=0.9\textwidth]{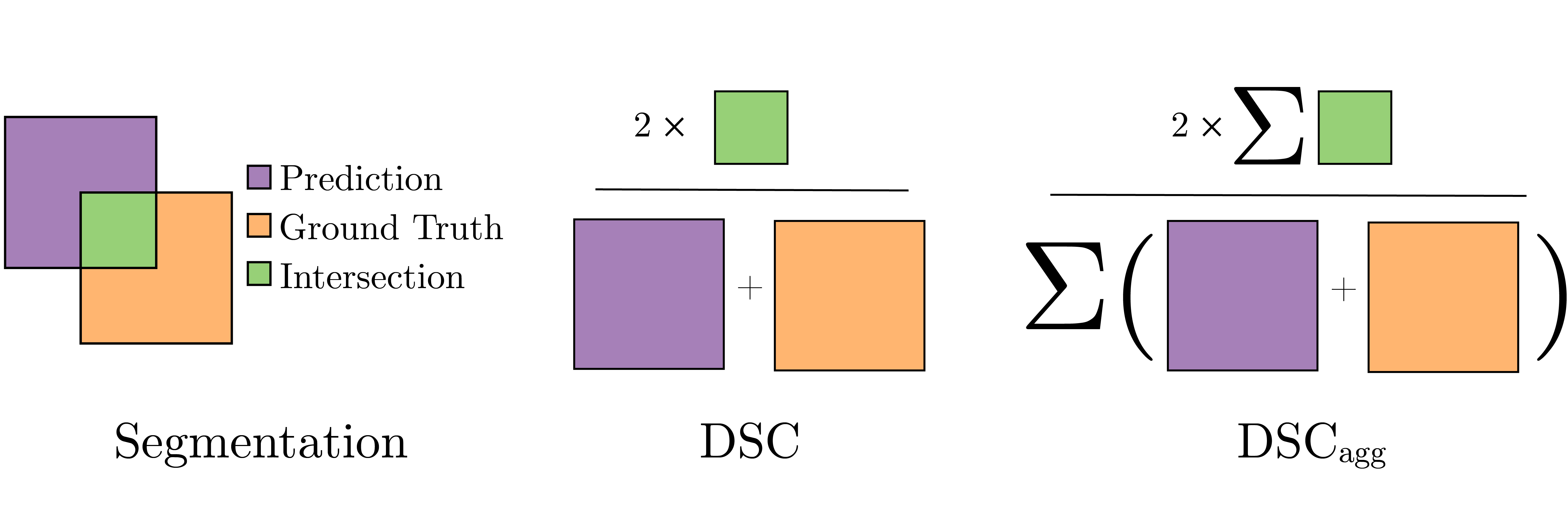}
\caption{%
Visual representation of the Dice Similarity Coefficient ($DSC$) and 
its aggregated version ($DSC_{agg}$).
Color marks different volumes, namely the expert annotated ground truth volume in orange, the predicted segmentation in purple, and the intersection volume indicating the true positive is green.
%
%
}
\label{fig4}
\end{figure}

The $DSC_{agg}$ is computed for both non-background classes GTVp and GTVn, aggregated over all test samples provided by the organizers. The final scoring is determined by the mean of the $DSC_{agg}$ over both types of GTV.

\subsection{Experiments}
We tested a custom framework on the data for Task 1 based on the MONAI package \cite{monai}, using the U-Net, 3D U-Net, and Swin UNETR architectures. 
The resulting training speed, validation results, and tinkering time were less optimal compared to the initial nnU-Net V2 experiments, which resulted in the decision to not further investigate the possibilities for this custom framework.

Furthermore, we also investigated on Task 1 the viability of using SAM \cite{medsam}, the foundation model (FM) pretrained on a large medical dataset. 
Since SAM needs user input in the form of clicks or bounding boxes, we adopted the Autogluon package that uses the SAM model as baseline \cite{agtabular,kirillov2023segment,tang2024autogluon}.
Zero-shot evaluation provided no realistic segmentation, therefore fine-tuning was required. For fine-tuning 3D image mask pairs were converted into 2D slices which served as input for the model.
The model was trained for up to 2 hours (6 epochs) on the Hàbròk HPC cluster, from which the optimal intermediate model was selected based on the validation Intersection over Union (IoU) score.
Since this method is a 2D approach, the individual slices were fed into the network for inference, stacking the output to create the volumetric segmentation. 
This preliminary test of the fine-tuned SAM approach resulted in a $DSC_{agg}$ of 0.70 on an internal test set.
Due to the more promising results obtained by nnU-Net, and the preference for 3D modalities \cite{unet_dimension_comparison}, we decided to not further investigate this approach for this challenge.

\begin{figure}[t]
\centering
\includegraphics[width=0.95\textwidth]{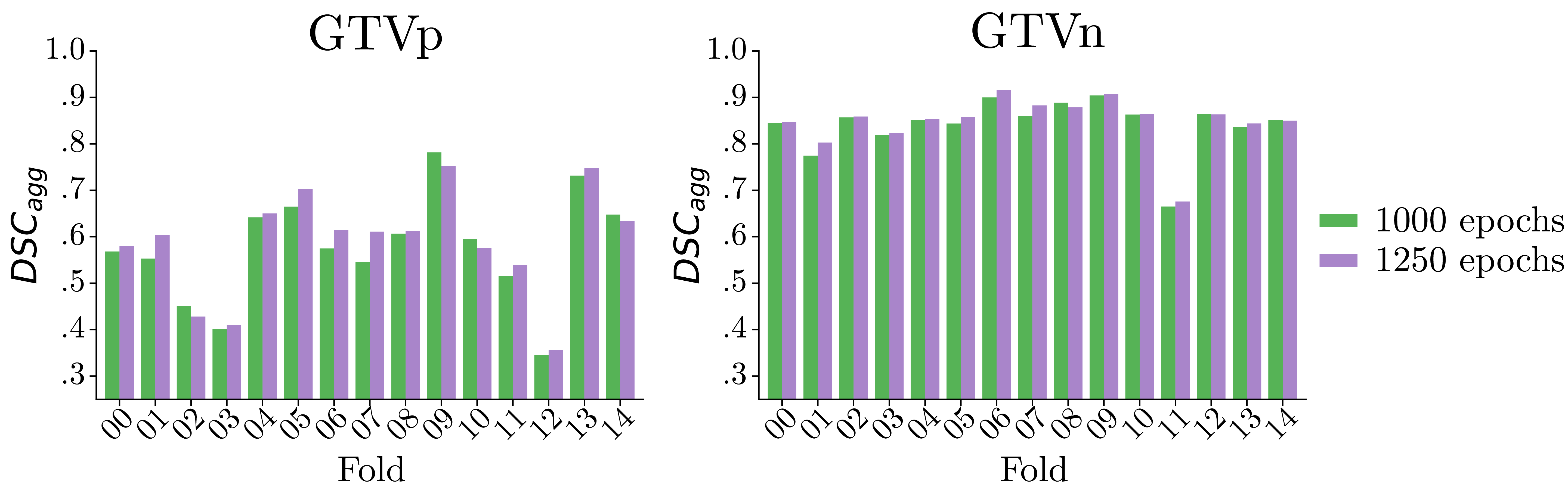}
\caption{Comparison of 15-fold cross validation of nnU-Net V2 using 1000 and 1250 epochs for mid-RT segmentation (Task 2). The validation $DSC_{agg}$  is displayed for each fold, displaying an increase in performance by training for 1250 epochs above 1000 epochs. The model trained for 1250 epochs yields increased performance in terms of $DSC_{agg}$ for 11 
and 12 
out of 15 folds for GTVp and GTVn, respectively. 
The $DSC_{agg}$ across all validation samples 
reached 
0.74 for 1250 epochs and 0.73 for 1000 epochs.}
\label{fig3}
\end{figure}

We adopted the nnU-Net V2 framework and increased the number of folds in the cross-validation ensemble from 5 to 15, to increase the robustness and variability of the ensemble. These benefits result from both an increase in the number of training samples per fold (140 instead of 120) and the use of a greater number of models trained on different subsets (15 instead of 5).
The resulting $DSC_{agg}$ scores on the validation sets were promising, leading us to adhere to this regime. 
For the first task, we explored various numbers of epochs, but this did not prompt us to deviate from the initial 1000 epochs set by the framework. 
For Task 2, we have investigated the impact of the number of epochs using 15-fold cross-validation and computing the $DSC_{agg}$ aggregated over the validation results over all folds.
The number of epochs used in our final method is 1250, based on the improved results compared to the 1000 epochs 15-fold cross-validation.
The increase in epochs resulted in a better $DSC_{agg}$ for 11 out of 15 folds for GTVp and 12 out of 15 folds for GTVn, as shown in Fig. \ref{fig3}. The final validation $DSC_{agg}$ computed over all samples increased from 0.73 for 1000 epochs to 0.74 for 1250 epochs. A comparison of training 5-folds and 15-folds resulted in a similar validation $DSC_{agg}$ of 0.74 for both models, both trained for 1250 epochs.
The source-code used for this study have been published on Zenodo, see the repository \cite{zenodo}.

\section{Results}
\subsection{Pre-RT segmentation}
The resulting method that is used for pre-RT segmentation of GTVp and GTVn (Task 1) is a 15-fold cross-validation ensemble of the nnU-Net V2 $3d_{fullres}$ configuration, in which each fold was trained on 280 training samples.
The 280 training samples consisted of 140 pre-RT and 140 mid-RT scans (from the same 140 patients, but treated as separate training inputs) with corresponding label masks. 
The model was trained over 1000 epochs. 
The increase of the number of ensembles from 5-fold cross-validation to 15-fold cross-validation ensembles increases the inference time, i.e. the number of forward passes through a model is increased from 5 to 15 for predicting a single segmentation.
Due to the inference time limit of the Docker image on the challenge platform of 20 minutes (NVIDIA T4 Tensor Core GPU), we decided to disable the test-time augmentation.
The reasoning is that increasing variability by increasing the number of folds has a larger impact than the data augmentation, in the form of flips in all dimensions, on the resulting segmentation quality.
%
The preliminary development phase, which consists of 2 patient samples, resulted in a mean $DSC_{agg}$ of 0.89. The final testing (on 50 patients) of the method resulted in a mean $DSC_{agg}$ of 0.77 for GTVp and 0.85 for GTVn. Resulting in a final mean $DSC_{agg}$ of 0.81.

\subsection{Mid-RT segmentation}
We applied a similar approach as in Task 1 for GTVp and GTVn segmentation on mid-RT data (Task 2), utilizing a 15-fold cross-validation ensemble of the nnU-Net V2 $3d_{fullres}$ configuration. 
In contrast to Task 1, the provided data includes more than just a single MRI volume, allowing for the use of multi-channel input.
The mid-RT MRI volume is complemented by the corresponding pre-RT MRI volume and label mask.
As a result, it is not possible to increase the number of initially provided samples when the pre-RT is added to the input.
The pre-RT data consisted of MRI volumes with corresponding segmentation masks, along with the registered versions aligned with the mid-RT volumes. 
We opted for including the registered pre-RT MRI volume (as the second channel) and the corresponding registered annotated mask for GTVp and GTVn (as the third channel) since the aligned data provides local information of the same region. Hence, the training of all folds was performed on 140 3-channels patient samples. Each fold was trained for 1250 epochs, and the resulting models were used for the final ensemble.
Again, we disabled test-time augmentation to reduce the inference time such that the inference time per sample falls within the 20-minute time limit.

To infer the model quality, we computed the $DSC_{agg}$ across all validation samples across all 15 folds, which results in a single $DSC_{agg}$ value across the validation test set. This resulted in a $DSC_{agg}$ of 0.62 for GTVp and 0.86 for GTVn, resulting in a final $DSC_{agg}$ of 0.74. The preliminary development phase, which consists of 2 patient samples, resulted in a mean $DSC_{agg}$ of 0.75. The final testing of the method (on data from 50 patients) resulted in a mean $DSC_{agg}$ of 0.54 for GTVp and 0.86 for GTVn. Resulting in a mean $DSC_{agg}$ of 0.70.

\section{Discussion} 
This study investigated the performance of several deep learning methods for segmenting GTVp and GTVn in pre-RT and mid-RT T2-weighted MRI scans of HNC patients. 
Initial experiments resulted in a final approach using the nnU-Net V2 framework \cite{nnUnet,nnunetv2}. This framework configures a U-Net architecture \cite{Unet}, , alongside the hyper-parameters and data processing steps, based on the provided training data. The nnU-Net framework has demonstrated high performance for Head and Neck Cancer segmentation in PET/CT \cite{nnUnet_2,nnUnet_1} and lymph node segmentation in MRI \cite{REINDERS2024100655}.
To increase the variability in the ensemble, and increase the number of training samples per fold, we adapted the framework to work with 15-fold cross-validation ensembles, instead of the original 5-fold cross-validation. 
For Task 1, the training data was augmented by incorporating the Mid-RT MRI volumes with corresponding label masks.
For Task 2, the increased complexity of the three-channel input (Mid-RT MRI, Registered pre-RT MRI and label mask) causes that more epochs were necessary to reach the optimal $DSC_{agg}$ scores in the validation set (1250 instead of 1000).
The final trained models yield convincing results in terms of the $DSC_{agg}$ values.

The final blind test on data from 50 patients resulted in a $DSC_{agg}$ of 0.81 for GTVp and GTVn segmentation on pre-RT MRI volumes (Task 1) and a $DSC_{agg}$ of 0.70 for GTVp and GTVn segmentation on mid-RT MRI volumes (Task 2).
Task 2 resulted in a lower score compared to Task 1, even though the provided information as input was increased.
The metastatic lymph nodes (GTVn) had similar results for both tasks in terms of $DSC_{agg}$, 0.85 for Task 1 and 0.86 for Task 2.
Consequently, the lower result is caused by a worse segmentation of the primary gross tumor volume (GTVp).
In Task 1, the GTVp resulted in a $DSC_{agg}$ of 0.77, followed by a lower $DSC_{agg}$ of 0.54 for Task 2.
The observed decrease in segmentation performance for GTVp from pre-RT to mid-RT could potentially be related to the large reduction in the size of the volumes (possibly due to effective RT). Furthermore, a decrease in the contrast results in less clear tumor boundaries. Future work aimed at improving segmentation performance for mid-RT GTVp  with lower contrasts holds significant potential. Additionally, advancements in the registration process may further improve performance for both labels in mid-RT segmentation.

MRI-based HNC tumor segmentation was less explored in previous studies compared to CT/PET, possibly due to the later inclusion of MRI into clinical radiotherapy flow of HNC.
However, with the increasing use of MRI in radiotherapy—such as employing MR-linac for visualization of the head and neck region before each radiotherapy fraction—automatic 
segmentation of the radiotherapy target volumes (GTVp and GTVn), as well as critical organs in the irradiated volume,
is potentially crucial to realize real-time accurate adaptive radiotherapy. 

This study has some limitations.
Firstly, a low sample size of 150 patients was used for training. To overcome these limitations, future work might benefit from an increase in sample size and utilize federated learning allowing to incorporate multi-center data to enhance the performance and generalization of the study.
Secondly, the advantage of MRI over CT/PET for HNC tumor segmentation remains to be proven.
In the future, a multi-center and multi-modality (CT/PET/MRI) dataset should be acquired to further evaluate the efficacy of nnU-Net for HNC tumor segmentation.

In conclusion, increasing the number of folds from 5 to 15 for the cross-validation ensemble displayed to have a positive effect on the segmentation performance using the nnU-Net V2 framework for GTVp and GTVn segmentation, in both pre-RT and mid-RT T2-weighted MRI volumes.

\begin{credits}
\subsubsection{\ackname} 
We thank the Center for Information Technology of the University of Groningen for their support and for providing access to the Hábrók high-performance computing cluster.

\subsubsection{\discintname}

The authors declare no competing interests.
\end{credits}
%
%
%
%
\bibliographystyle{splncs04}
\bibliography{references}

\end{document}